# Entanglement based frequency-time coding quantum key distribution

**Bing Qi**
*Center for Quantum Information and Quantum Control, University of Toronto, Toronto, Ontario, Canada*
*Dept. of Electrical and Computer Engineering, University of Toronto, Toronto, Ontario, Canada*
*bing.qi@utoronto.ca*

**Abstract:** We extend the prepare-and-measure frequency-time coding quantum key distribution (FT-QKD) protocol to an entanglement based FT-QKD protocol. The latter can be implemented with a correlated frequency measurement scheme based on a time resolving single photon detector.

## 1. Introduction

One important practical application of quantum information is quantum key distribution (QKD) [1]. Most practical QKD systems are based on either polarization coding or phase coding. Unfortunately, these two coding schemes suffer from polarization and phase instabilities in optical fiber induced by environmental noise. On the contrary, the frequency-time coding QKD (FT-QKD) scheme proposed in [2] is intrinsically insensitive to the polarization and phase fluctuations. This suggests the FT-QKD could be a more robust solution in practice.

As shown in Fig.1, in the prepare-and-measure FT-QKD, Alice randomly chooses to use either "frequency-basis" or "time-basis" to encode her random bits. In the frequency-basis, information is encoded on the central frequency of a single-photon pulse which has a very small line-width. In the time-basis, information is encoded on the time delay (defined relatively to a synchronization pulse) of a single-photon pulse which has a very small temporal duration. Upon receiving Alice's photon, Bob randomly chooses to measure either its frequency or its arrival time. After the quantum transmission stage, Alice and Bob compare their bases through a public authenticated channel and they only keep the results when they happen to use the same basis. Given the conditional variance of Bob's measurement results is below certain threshold, they can further generate identical secure key by performing error correction and privacy amplification. The security of the FT-QKD protocol can be intuitively understood from the energy-time uncertainty relation, which puts a constraint on Eve's ability to simultaneously determine both the frequency and the arrival time of a photon.

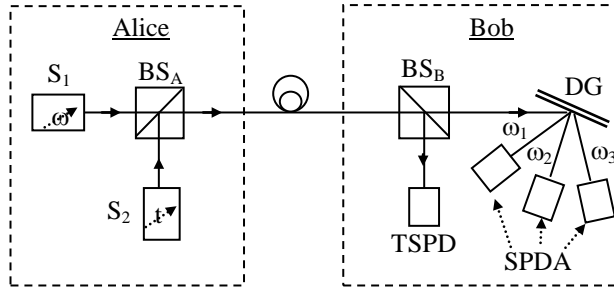

Fig.1 Schematic diagram of the prepare-and-measure FT-QKD [2]: $S_1$-narrowband frequency tunable single photon source; $S_2$-broadband single photon source with tunable time-delay; $BS_A/BS_B$-beam splitters; TSPD-time-resolving single photon detector; DG-dispersive grating; SPDA-single photon detector array.

The FT-QKD protocol is interesting in principle. However the system proposed in [2] is too complicated to be attractive in practice. Here, we extend the FT-QKD to an entanglement based scheme and discuss the feasibility of implementing the FT-QKD with today's technology.

## 2. The entanglement based FT-QKD protocol

The prepare-and-measure FT-QKD protocol shown in Fig.1 can be extended into an entanglement based QKD protocol, as shown in Fig.2. A source generating energy-time entangled photon pairs can be placed either at Alice's station or between Alice and Bob. The energy and time of the two photons in the same pair are Einstein-Podolsky-Rosen (EPR) [3] correlated. One photon from an EPR pair is sent to Alice and the other one is sent to Bob. Passively determined by a beam splitter, Alice (Bob) randomly measures either the arrival time or the frequency (wavelength) of each incoming photon. After the quantum transmission stage, Alice and Bob compare their measurement bases for each photon pair and only keep the results when they happen to use the same basis. Given the conditional variance is below certain threshold, they can further distillate out secure key.

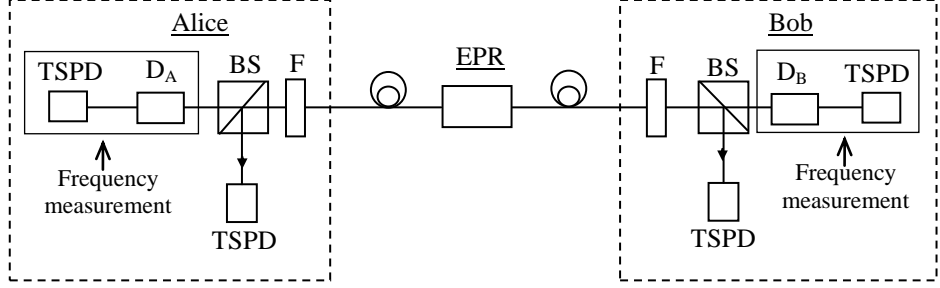

Fig.2 Schematic diagram of the entanglement based FT-QKD: EPR-frequency-time entangled source; BS-beam splitter; F-spectral and temporal filters; $D_A$-dispersive component with positive dispersion coefficient; $D_B$-dispersive component with negative dispersion coefficient; TSPD-time-resolving single photon detector.

*2.1 Spontaneous parametric down-conversion (SPDC) source*

In practice, the above energy-time entangled photon pairs can be generated through nonlinear optical processes, such as spontaneous parametric down-conversion (SPDC). In this process, a pump photon (with a central frequency of $\nu_P$) spontaneously decays into a pair of daughter photons ($\nu_A$ and $\nu_B$) in a nonlinear crystal. The conservation of energy and momentum implies that the generated daughter photons are entangled in spectral and spatial domains. Each individual photon of an EPR pair has both a broad spectral bandwidth and a large temporal width. This suggests that when Alice and Bob perform time or frequency measurement, individually, they will observe a large uncertainty in their measurement results. However, as long as they use the same basis, their measurement results are highly correlated, i.e. $\nu_A + \nu_B \cong \nu_P$ and $t_A \cong t_B$.

*2.2 A correlated frequency measurement scheme of entangled photon pair*

Instead of using an SPD array to measure the frequency of the incoming photon (as shown in Fig.1), Alice and Bob can use a highly dispersive element followed by a time resolving SPD (as shown in Fig.2). The dispersive element introduces a frequency-dependent time delay, thus information encoded in spectral domain will be transferred into time domain and can be decoded by measuring its arrival time. However, this idea cannot be applied directly. This is because each individual photon has a large intrinsic time uncertainty and the SPD cannot distinguish it from the frequency-dependent time delay.

Fortunately, the two photons in an EPR pair are entangled in both spectral and time domain, so the intrinsic time uncertainty of each individual photon can be cancelled out. In Fig.2, the dispersion coefficients of the two dispersive elements are chosen to satisfy $D_B = -D_A$. By using a suitable time reference, the detection time $T_A$ of Alice's SPD in frequency-basis is given by

$$T_A = t_A + D_A(\nu_A - \nu_0) \tag{1}$$

Similarly

$$T_B = t_B + D_B(\nu_B - \nu_0) \tag{2}$$

From $\nu_A + \nu_B \cong \nu_P$, $t_A \cong t_B$ and $D_B = -D_A$, it is easy to show that $T_A$ and $T_B$ are highly correlated.

## 3. Simulation results

The security of the FT-QKD protocol could be established by connecting it to the squeezed state QKD protocol [4], whose security against the most general attack has been proven by Gottesman and Preskill [5]. This is because both protocols are based on the uncertainty relation between a pair of observables. One nice feature of Gottesman-Preskill's proof is that both the BB84 QKD protocol and the squeezed state QKD protocol (thus the FT-QKD protocol) are studied under the same scope. This allows us to apply many important results developed in the BB84 QKD, such as decoy state idea and the squash model of threshold detector, into the FT-QKD protocol.

The secure key rate is given by [5]:

$$R = \tfrac{1}{2} Q_1 [1 - f(e) H_2(e) - H_2(e)] \tag{3}$$

In (3), *e* is the observed quantum bit error rate (QBER), $Q_1$ is the overall gain, *f(x)* is the bidirectional error correction efficiency, and $H_2(x)$ is the binary entropy function. The QBER is further determined by $e \leq \frac{2\Delta}{\pi}\exp(\frac{-\pi}{4\Delta^2})$, where $\Delta^2$ is a measure of the conditional variance.

The conditional variance (thus the intrinsic QBER) of the FT-QKD is mainly determined by the finite temporal and spectral resolutions of the detection system. Specifically, in the entanglement FT-QKD scheme shown in Fig.2, given the dispersion coefficient of the dispersive elements, the intrinsic QBER is mainly determined by the time jitter of time resolving SPDs. A commercial dispersion compensation module based on fiber Bragg grating (FBG) technology can provide a dispersion coefficient as larger as 7000ps/nm with a moderate loss of 5dB (www.teraxion.com). If the time resolution of the SPD is 50ps, then the spectral resolution will be about 7pm.

We calculate the intrinsic QBER as a function of the time resolution of the SPD. Here, we assume that the QKD system is operated at telecom wavelength and the dispersion coefficient D = 7000ps/nm. The simulation results are show in Fig.3: the QBER is about 5% for a time jitter of 70ps. The time jitter of a state-of-the-art superconducting nanowire SPD (SNSPD) can be as small as 40ps [6], and the resulting intrinsic QBER is about 0.05%.

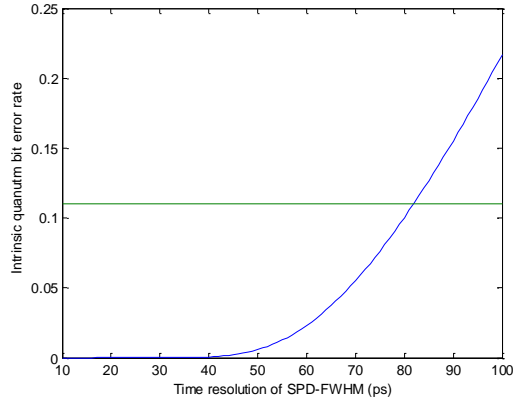

Fig.3 (Simulation results) Intrinsic quantum bit error rate of the entanglement FT-QKD protocol. Here we assume D =7000ps/nm. The QBER is about 5% for a time jitter of 70ps. We also show the 11% security bound.

## 4. Conclusion

One major advantage of the FT-QKD protocol is the robust against environmental noise: the frequency/time coding scheme is intrinsically insensitive to the polarization and phase fluctuations. This could improve the stability of a practical QKD system dramatically. We extend the prepare-and-measure FT-QKD protocol to an entanglement based FT-QKD protocol which is more appealing in practice. Furthermore, we propose a correlated frequency measurement scheme by using time resolving SPDs. Simulation results show the feasibility of the FT-QKD protocol.

**Acknowledgement**: The author is very grateful to Hoi-Kwong Lo and Li Qian for their support and helpful comments. The author also thanks John Sipe, Eric Chitambar, Christian Weedbrook, Wolfram Helwig, Wei Cui, Luke Helt, and Sergei Zhukovsky for helpful discussions. Financial support from CFI, CIPI, the CRC program, CIFAR, MITACS, NSERC, OIT, and QuantumWorks is gratefully acknowledged.